\documentclass[journal]{IEEEtran}

\usepackage{algorithmic}
\usepackage{textcomp}
\def\BibTeX{{\rm B\kern-.05em{\sc i\kern-.025em b}\kern-.08em
    T\kern-.1667em\lower.7ex\hbox{E}\kern-.125emX}}

\usepackage{graphicx}
\usepackage[font=footnotesize]{subfig}
\usepackage{float}
\usepackage{color}
\usepackage{bbm}
\usepackage{dsfont}
\usepackage{cite}
\usepackage[hidelinks, bookmarks=false, pdfkeywords={physical layer security; PLS; SOP; reconfigurable intelligent surfaces; RIS; vehicular communications; vehicular network; V2V; V2I }]{hyperref}
\usepackage{balance}
\usepackage{url}
\usepackage{times}
\usepackage{amsbsy}
\usepackage[T1]{fontenc}
\usepackage[latin9]{inputenc}
\usepackage{tablefootnote}
\usepackage{threeparttable}
\usepackage{amsmath,amssymb,amsfonts}
\usepackage{subfig}
\usepackage[font=small]{caption}
\usepackage{multirow}
\renewcommand\IEEEkeywordsname{Index Terms}

\usepackage{resizegather}

\usepackage{suffix}
\usepackage{mathtools}

\DeclarePairedDelimiterX\MeijerM[3]{\lparen}{\rparen}%
{\,#3\delimsize\vert\begin{smallmatrix}#1 \\ #2\end{smallmatrix}}

\newcommand\MeijerG[8][]{%
  G^{\,#2,#3}_{#4,#5}\MeijerM[#1]{#6}{#7}{#8}}

\WithSuffix\newcommand\MeijerG*[7]{%
  G^{\,#1,#2}_{#3,#4}\MeijerM*{#5}{#6}{#7}}

\newcommand{\subparagraph}{}
\usepackage{titlesec}

\titlespacing*{\subsubsection}
{0pt}{.086cm}{.1cm}

\usepackage{romannum}

\begin{document}

\bstctlcite{IEEEexample:BSTcontrol}

\title{Secure Vehicular Communications through Reconfigurable Intelligent Surfaces}

\author{
Yun Ai, Felipe A. P. de Figueiredo, Long Kong,  Michael Cheffena,\\ Symeon Chatzinotas, \IEEEmembership{Senior Member, IEEE}, and Bj{\"o}rn Ottersten, \IEEEmembership{Fellow, IEEE}

\thanks{Y. Ai and M. Cheffena are with the Norwegian University of Science and Technology (NTNU), F. Figueiredo is with Instituto Nacional de Telecomunicações (INATEL), Brazil, L. Kong, S. Chatzinotas, and B. Ottersten are with the University of Luxembourg.}
}

%

{}


\maketitle

\begin{abstract}
Reconfigurable intelligent surfaces (RIS) is considered as a revolutionary technique to improve the wireless system performance by reconfiguring the radio wave propagation environment artificially. Motivated by the potential of RIS in vehicular networks, we analyze the secrecy outage performance of RIS-aided vehicular communications in this paper. More specifically, two vehicular communication scenarios are considered, i.e., a vehicular-to-vehicular (V2V) communication where the RIS acts as a relay and a vehicular-to-infrastructure (V2I) scenario where the RIS functions as the receiver. In both scenarios, a passive eavesdropper is present attempting to retrieve the transmitted information. Closed-form expressions for the secrecy outage probability (SOP) are derived and verified. The results demonstrate the potential of improving secrecy with the aid of RIS under both V2V and V2I communications.
\end{abstract}


\IEEEpeerreviewmaketitle

\section{Introduction \label{sec:introduction}}

\IEEEPARstart{R}{econfigurable} intelligent surfaces (RIS) have recently appeared as a revolutionary technique to enhance network coverage and overcome the high attenuation of millimeter wave (mmWave) and THz systems \cite{wu2019towards}. By intelligently controlling a large number of low-cost passive reflecting elements, the electromagnetic waves can be adapted to the propagation environment. Thereby, the RIS functions as a reconfigurable lens or reconfigurable mirror to beamform the transmitted signals towards the desired user \cite{basar2019wireless}. RIS are also widely known as intelligent reconfigurable surfaces (IRS), software-controllable surfaces, digitally controllable scatterers, and large intelligent surfaces in literatures \cite{wu2019towards, basar2019wireless}.


Physical layer security (PLS) is widely considered as a complement to conventional application layer encryption techniques to enhance the communication secrecy in future (5G and beyond) communication systems \cite{barros2006secrecy}. It has been demonstrated both theoretically and experimentally that channel fading, which is usually regarded as an adverse factor in terms of reliability, can be utilized to enhance communication security against eavesdropping \cite{barros2006secrecy, ai2019secrecyisl, qiao2020secure}. Due to potential of RIS and PLS technologies in future networks, investigation of the combination of PLS and RIS-assisted systems has attracted attention from the community recently \cite{qiao2020secure, yang2020secrecy, makarfi2020physical}. The optimized beamforming and phase shift design for secrecy rate of an RIS-assisted mmWave system is investigated in \cite{qiao2020secure}. In \cite{yang2020secrecy}, the secrecy outage probability (SOP) of an RIS-aided system is studied while the authors in \cite{yang2020secrecy} focus on the SOP of an RIS-aided non-orthogonal multiple access (NOMA) system. The expressions, in integral form, for the average secrecy capacity (ASC) of an RIS-assisted vehicular network are derived in \cite{makarfi2020physical}.

\begin{figure}[t!]
\centering
  \includegraphics[width=\linewidth,keepaspectratio,angle=0]{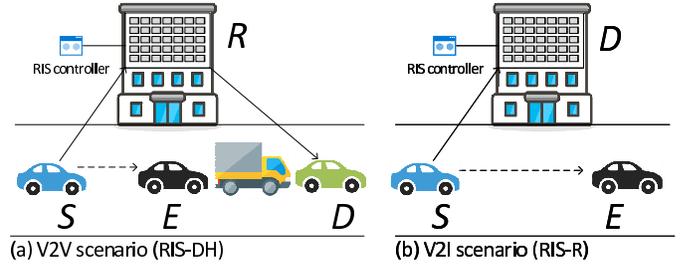}
  \caption{Considered PLS scenarios for RIS-aided vehicular networks.}
  \label{fig:system_model}
\end{figure}

The realization of future autonomous vehicles requires robust connections and high quality-of-service (QoS) between vehicles (i.e., vehicular-to-vehicular (V2V)) as well as between vehicle and infrastructure (i.e., vehicular-to-infrastructure (V2I)) \cite{ai2018physical}. The aforementioned advantages of RIS technique make RIS-assisted vehicular communication an appealing option to enhance vehicular network connectivity \cite{liu2020reconfigurable}. Motivated by the latest advances in PLS analysis of RIS-assisted systems as well as the potential of RIS-assisted communication in vehicular networks, we study herein the secrecy performance of RIS-assisted vehicular communications under passive eavesdropping. More specifically, we consider the SOP performance under two communication scenarios of vehicular communication, i.e., V2V and V2I scenarios. In the V2V scenario, the RIS assists two vehicles that are blocked by other objects to communicate with high QoS. In the V2I scenario, the vehicle sends essential information to the RIS that is close to the receiver to ensure robust transmission of important messages to the intelligent transport infrastructure.


The main contributions of this paper are: (\romannum{1}) We analyze the secrecy performance of RIS-assisted vehicular communication under two realistic cases, where RIS are used as part of dual-hop system and part of receiver, respectively; (\romannum{2}) By avoiding applying the central limit theorem (CLT) and instead adopting a more versatile approach in the RIS analysis, the obtained results are also valid when the number of RIS elements is small; and (\romannum{3}) We present some accurate or exact statistics for RIS related signal-to-noise ratios (SNRs) (in Propositions 1, 2 and 3), which can be useful for RIS-related analysis.




\emph{Notations}: $[0,x]^{+} = \max(x, 0)$, $\mathbb{E}[\cdot]$ is the expectation operator, $\Gamma(\cdot)$ and $\Gamma(\cdot, \cdot)$ are Gamma and incomplete Gamma functions, respectively \cite[Eq.~8.3]{jeffrey2007table}, $K_{v}(\cdot)$ is the modified Bessel function of second kind with order $v$ \cite[Eq.~8.407]{jeffrey2007table}, $J_{v}(\cdot)$ is the Bessel function of first kind \cite[Eq.~8.402]{jeffrey2007table}, $G^{m,n}_{p,q}\!(\cdot)$ is the Meijer G-function \cite[Eq.~(9.3)]{jeffrey2007table}, $H^{m,n}_{p,q}(\cdot)$ is the Fox H-function \cite[Eq.~1.2]{mathai2009h}, and $H^{m,n:s,t:i,j}_{p,q:u,v:e,f}(\cdot)$ is the extended generalized bivariate Fox H-function \cite[Eq.~2.56]{mathai2009h}.




\section{Channel and System Models \label{sec:channel_system_model}}

In this paper, we consider the secrecy outage probability of RIS-assisted V2V and V2I systems. The considered classic Wyner's wiretap model is illustrated in Fig. \ref{fig:system_model}.

\subsection{V2V Communications \label{subsec:v2v_system_model}}

In the V2V case, a vehicle \emph{S} communicates secret information with another vehicle \emph{D} that has blockage between them with the aid of an RIS with $N$ elements. The signals sent by \emph{S} is overheard by an eavesdropper \emph{E} close to \emph{S}. All vehicles are assumed to be equipped with single antennas for simplicity.

The received signal at the receiver vehicle \emph{D} via the RIS is
\begin{align}
y_{D}  =   \sqrt{ P_{s} } \cdot \mathbf{h}_{SR}^{T} \mathbf{\omega}  \mathbf{h}_{RD}    \cdot s + w_{0} ,
\label{eq:v2v_reced_signal}
\end{align}
where $P_{s}$ is the transmit power of $S$, $s$ is the transmitted signal with unit energy, $w_{0}$ is the zero-mean additive white Gaussian noise (AWGN) with variance $N_{0}$, $\mathbf{\omega} = \mathrm{diag}(\varpi_{1}(\phi_{1})e^{j\phi_{1}}, \dots, \varpi_{N}(\phi_{N})e^{j\phi_{N}})$ is the diagonal matrix consisting of the reflection coefficients produced by each reflection element of the RIS. The vector $\mathbf{h}_{SR}$ contains the channel gains from \emph{S} to each element of RIS and the vector $\mathbf{h}_{RD}$ includes the channel gains from each element of RIS to \emph{D}, which are expressed as \cite{yang2020performance}
\begin{subequations}\label{eq:v2v_channel_gain}
\begin{eqnarray}
\mathbf{h}_{SR} & = & \mathbf{\alpha}^{T} \Theta \cdot d_{SR}^{ - \frac{ p_{1} }{ 2 } } , \\
\mathbf{h}_{RD} & = & \mathbf{\beta}^{T}  \mathbf{\varphi} \cdot d_{RD}^{ - \frac{ p_{1} }{ 2 } },
\end{eqnarray}
\end{subequations}
where the column vectors $\mathbf{\alpha}$ and $\mathbf{\beta}$ contain the amplitudes of the corresponding channel gains. Each element of $\mathbf{\alpha}$ follows independent Rayleigh distribution resulting from the scattering around the vehicle. Similarly, every element of $\mathbf{\beta}$ is also independent Rayleigh distributed. $\Theta = [e^{ -j \theta_{1}}, \cdots,  e^{ -j \theta_{N}}]$ and $\mathbf{\varphi} = [e^{ -j \varphi_{1}}, \cdots,  e^{ -j \varphi_{N}}]$ with $\theta_{n}$ and $\varphi_{n}$, $n = 1, \cdots, N$, being phase of the corresponding link; $d_{xy}$ is the distance between nodes \emph{x} and \emph{y}; and $p_{1}$ is the path loss exponent for the link from or to the RIS.


From (\ref{eq:v2v_reced_signal}) and (\ref{eq:v2v_channel_gain}), the instantaneous SNR $\gamma_{\mathrm{\emph{D}}}$ at \emph{D} becomes
\begin{align}
\gamma_{\mathrm{\emph{D}}} = \frac{ P_{s} \! \cdot \! \biggl|   \sum\limits_{n = 1}^{N} \alpha_{n} \beta_{n} \varpi_{n}(\phi_{n}) \cdot e^{j( \phi_{n} - \theta_{n} - \varphi_{n} )}  \biggr|^{2} }{  N_{0} d_{SR}^{ p_{1} }  d_{RD}^{ p_{1} } } .
\label{eq:v2v_reced_snr}
\end{align}


We first consider perfect knowledge of the channel state information (CSI) at RIS as in \cite{qiao2020secure, yang2020secrecy, makarfi2020physical}, which enables ideal phase shifting (i.e., $\varpi_{n}(\phi_{n}) = 1$ and $\sigma_{n}^{D} = \phi_{n} - (\theta_{n} + \varphi_{n}$) = 0). Then, the maximum instantaneous SNR can be achieved at \emph{D} and is expressed as
\begin{align}
\gamma_{D} = \overline{\gamma}_{D} \cdot  \left(   \sum\limits_{n = 1}^{N} \alpha_{n} \beta_{n}  \right)^{2} = \overline{\gamma}_{D} \cdot  \left(   \sum\limits_{n = 1}^{N}    \emph{a}_{n} \right)^{2} = \mathcal{A}^{2},
\label{eq:v2v_max_snr_d}
\end{align}
where $ \overline{\gamma}_{D} \! = \!\! \frac{ P_{s} }{ N_{0} d_{SR}^{ p_{1} } d_{RD}^{ p_{1} } } $, $\emph{a}_{n} \! = \!\! \alpha_{n} \beta_{n}$, and $\mathcal{A}\! = \!\! \sqrt{\overline{\gamma}_{D} }  \sum_{n = 1}^{N} \! \emph{a}_{n}$.

When $N$ is large, the CLT can be applied and the random variable (RV) $\mathcal{A}$ can be approximated by a Gaussian RV and the RV $\gamma_{D}$ can be considered to follow noncentral-$\chi^{2}$ distribution \cite{basar2019wireless}. Nevertheless, the CLT approximation becomes inaccurate while $N$ is not large. In \cite{atapattu2020reconfigurable}, an approximation of $\gamma_{D}$ is obtained by considering the RV $\emph{a}_{n} = \alpha_{n} \beta_{n}$ as Gamma distributed with $\alpha_{n}$ and $\beta_{n}$ being independent and identically distributed (i.i.d.) RVs. However, it is unrealistic to assume that $\alpha_{n}$ and $\beta_{n}$ are identically distributed since they correspond to two completely different propagation links. Here, we present in Proposition 1 the statistics of $\gamma_{D}$ under the assumption of $\alpha_{n}$ and $\beta_{n}$ being independent but not identically distributed (i.n.i.d.) RVs by approximating the RV $\mathcal{A}$ as a Gamma RV.



\textbf{Proposition 1}: When $h_{SR,n}  \sim  \mathcal{CN}(0, \nu_{SR} )$ and $h_{RD,n} \sim \mathcal{CN}(0, \nu_{RD} )$ are i.n.i.d. complex Gaussian RVs with $|h_{SR,n}|=\alpha_{n}$ and $|h_{RD,n}| = \beta_{n}$, then the RV $\gamma_{D} = \overline{\gamma}_{D} \cdot  \bigl(   \sum_{n = 1}^{N} \alpha_{n} \beta_{n}  \bigr)^{2}$ under both small and large values of $N$ can be accurately described by the following probability density function (PDF) and cumulative distribution function (CDF):
\begin{align}
f_{\gamma_{D}}(x)  = & \frac{ 1 }{ 2 \Gamma(k_{D}) \eta_{D}^{k_{D}} }  \cdot x^{ \left( \frac{ k_{D} - 2 }{ 2 } \right)} \cdot \exp\!\!\left( - \frac{ \sqrt{ x } }{ \eta_{D} } \right) ,
\label{eq:v2v_pdf_max_snr_d} \\
F_{\gamma_{D}}(x)  = & 1 - \frac{ 1 }{ \Gamma( k_{D} ) } \cdot \Gamma\!\left( k_{D}, \frac{ \sqrt{x} }{ \eta_{D} } \right) ,
\label{eq:v2v_cdf_max_snr_d}
\end{align}
where $k_{D}  =  \frac{ N \pi^{2} }{ 16 - \pi^{2}  } , \label{eq:v2v_statistics_aux1a}$ and $\eta_{D} = \frac{ \sqrt{ \overline{\gamma}_{D} } ( 16 - \pi^{2}  ) \sqrt{\nu_{SR} \nu_{RD} } } { 4 \pi }$.

\begin{figure}[t!]
\centering
  \includegraphics[width=0.96\linewidth,keepaspectratio,angle=0]{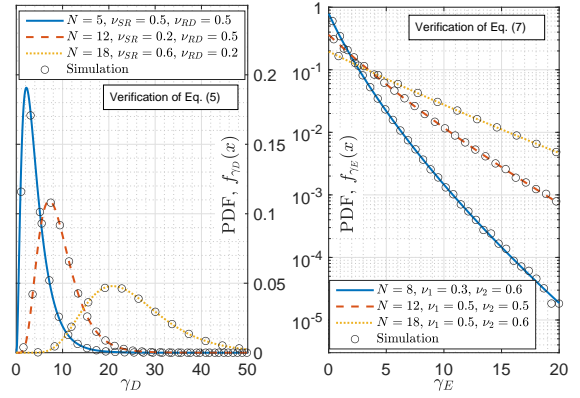}
  \caption{Verifications of statistics functions in Eqs. (\ref{eq:v2v_pdf_max_snr_d}) and (\ref{eq:v2v_pdf_snr_d2}).}
  \label{fig:statistics_verification}
   \vspace*{-3pt}
\end{figure}

\emph{Proof}: Please refer to Appendix A\ref{app_sec:proposition1} and Fig. \ref{fig:statistics_verification}.





Due to the high mobility of vehicles \emph{S} and \emph{D}, perfect phase estimation required for ideal RIS might be challenging. Next, we consider the worst case of RIS phase shifting, where the phase errors $\sigma_{n}^{D}$ are uniformly distributed in complex plane to evaluate the impact of imperfect RIS on secrecy performance. To obtain the statistics of RV $\gamma_{D}  =   \overline{\gamma}_{D} \cdot  \left|   \sum_{n = 1}^{N} \alpha_{n} \beta_{n} e^{j \sigma_{n}^{D}}   \right|^{2} $, we interpret the mathematical problem as an isotropic two-dimensional random walk, where the $n$-th step size is $\alpha_{n}\beta_{n}$ and $n$-th direction is $\sigma_{n}^{D}$ that is uniformly distributed.



\textbf{Proposition 2}: When $h_{SR,n} \sim \mathcal{CN}(0, \nu_{SR} )$ and $h_{RD,n} \sim \mathcal{CN}(0, \nu_{RD} )$ are i.n.i.d. complex Gaussian RVs with $|h_{SR,n}|=\alpha_{n}$ and $|h_{RD,n}| = \beta_{n}$, and $\sigma_{n}^{D}$ is uniformly distributed between $0$ and $2\pi$, the exact PDF and CDF of the RV $\gamma_{D} = \overline{\gamma}_{D} \cdot  \bigl|   \sum_{n = 1}^{N} \alpha_{n} \beta_{n} e^{j \sigma_{n}^{D}}   \bigr|^{2}$ are given by
\begin{align}
f_{\gamma_{D}}(x)  = & \frac{ 2 \cdot x^{ \frac{N-1}{2} } }{ \mathcal{B}  }   \cdot K_{N-1}\!\!\left( 2 \sqrt{ \frac{ x }{ \overline{\gamma}_{D} \nu_{SR} \nu_{RD} } } \right) , \label{eq:v2v_pdf_snr_d2} \\
F_{\gamma_{D}}(x)  = & \frac{  x^{ \frac{ N + 1 }{ 2 } }  }{ \mathcal{B}  }  \cdot  G^{2,1}_{1,3}\!\!\left( \!  \frac{ x }{ \overline{\gamma}_{D} \nu_{SR} \nu_{RD}  }   \left\vert^{  \frac{ 1 - N }{2}  }_{ \frac{N-1}{2}, -\frac{N-1}{2}, - \frac{ N+1 }{2}  } \right. \!\!\! \right) ,
\label{eq:v2v_cdf_snr_d2}
\end{align}
where $\mathcal{B} =  \Gamma(N)  \cdot ( \overline{\gamma}_{D} \nu_{SR} \nu_{RD} )^{  \frac{ N+1 }{ 2 } } $.

\emph{Proof}: Please refer to Appendix B\ref{app_sec:proposition2} and Fig. \ref{fig:statistics_verification}.



For the link between vehicles \emph{S} and \emph{E}, the double-bounce scattering components caused by scatterers around both vehicles' local environments lead to a cascaded Rayleigh fading process \cite{ai2018physical}. Therefore, we use the double Rayleigh model to characterize of the dynamic fading link between \emph{S} and \emph{E}. The PDF and CDF of the instantaneous SNR $\gamma_{E}$ are expressed as
\begin{align}
f_{\gamma_{E}}( x ) & =  \frac{ 2 }{ \overline{\gamma}_{E} }  \cdot  K_{0}\!\biggl( 2 \sqrt{  \frac{ x }{ \overline{\gamma}_{E} }  } \biggr), \label{eq:v2i_pdf_snr_e} \\
F_{\gamma_{E}}( x ) & =  1   -   2 \sqrt{  \frac{ x }{ \overline{\gamma}_{E} }  } \cdot  K_{1}\!\biggl( 2 \sqrt{  \frac{ x }{ \overline{\gamma}_{E} }  } \biggr),
\label{eq:v2i_cdf_snr_e}
\end{align}
where $\overline{\gamma}_{E} = \frac{ P_{s} }{ N_{0} d_{SE}^{ p_{2} } }$, and $p_{2}$ is the path loss exponent for the links between vehicles.


\subsection{V2I Communications \label{subsec:v2i_system_model}}

Under the V2I scenario, a vehicle \emph{S} sends essential information to the intelligent transportation infrastructure $D$ while an eavesdropper \emph{E} close to \emph{S} attempts to eavesdrop the signals sent by \emph{S}. The infrastructure $D$ consists of an RIS and RF receiver, where the RIS is deployed close to the RF receiver such that the channel attenuation between them can be ignored \cite{basar2019wireless}. Therefore, the RIS and RF receiver together are considered as a receiver from the perspective of analytical analysis.


Under the described V2I scenario with ideal phase shifting, the instantaneous SNR at \emph{D} can be written as
\begin{align}
\gamma_{D}  = \overline{\gamma}_{D} \cdot  \left(   \sum\limits_{n = 1}^{N} \alpha_{n}   \right)^{2} = \overline{\gamma}_{D} \cdot \mathcal{C}^{2} ,
\label{eq:v2i_snr_d}
\end{align}
where $ \overline{\gamma}_{D} = \frac{ P_{s} }{ N_{0} d_{SD}^{ p_{1} } } $ and $\mathcal{C} = \sum_{n = 1}^{N} \alpha_{n}$. To solve the statistics of $\mathcal{C}$ is equivalent to obtain the statistics of the received signal for the single-input multiple-output (SIMO) system with equal gain combining, where the exact closed-form solution is unavailable for $N > 2$. Next, we present an accurate approximation of $\gamma_{D}$ in Proposition 3.


\textbf{Proposition 3}: When $h_{SD,n} \sim \mathcal{CN}(0, \nu_{SD} )$ and $\alpha_{n} = |h_{SD,n}|$, the PDF and CDF of the RV $\gamma_{D} = \overline{\gamma}_{D} \cdot  \bigl|   \sum_{n = 1}^{N} \alpha_{n}  \bigr|^{2}$ can be closely approximated by
\begin{align}
f_{\gamma_{D}}(x)  = & \frac{ x^{N-1} }{ ( \overline{\gamma}_{D}\nu_{SD})^{N} \Omega_{D}^{N} \Gamma(N) }  \cdot \exp\!\left( - \frac{ x }{ \overline{\gamma}_{D} \nu_{SD} \Omega_{D} } \right) , \label{eq:v2i_pdf_snr_d}  \\
F_{\gamma_{D}}(x)  = & 1 - \frac{ 1 }{ \Gamma(N) } \cdot \Gamma\!\left( N, \frac{ x }{ \overline{\gamma}_{D} \nu_{SD} \Omega_{D} }  \right) ,
\label{eq:v2i_cdf_snr_d}
\end{align}
where $\Omega_{D} = 1+  \left[ \Gamma( \frac{3}{2} ) \right]^{2} \cdot (N-1)$.

\emph{Proof}: The results follow by employing the result in \cite[Eq.~(22)]{hadzi2009accurate} for the special case of independent Rayleigh RVs.



Under the V2I scenario, the PDF and CDF of the eavesdropper's SNR are given as in (\ref{eq:v2i_pdf_snr_e}) and (\ref{eq:v2i_cdf_snr_e}).

\section{Secrecy Performance Analysis \label{sec:secrecy_analysis}}

\subsection{Secrecy Outage Probability (SOP) \label{subsec:sop_analysis}}

The secrecy rate indicates the maximum achievable rate the main channel can achieve in secrecy. The instantaneous secrecy rate $C_{s}$ of the considered wiretap model is \cite{ai2020secrecy}
\begin{align}
C_{s}(\gamma_{D}, \gamma_{E}) = \left[\ln(1+\gamma_{D}) - \ln(1+\gamma_{E}), 0 \right]^{+},
\label{eq:inst_asc_def}
\end{align}
where $\gamma_{D}$ and $\gamma_{E}$ are the instantaneous SNRs of the main link from $S$ to $D$ and wiretap channel from $S$ to $E$, respectively.

Under passive eavesdropping, the legitimate transmitter $S$ and receiver $D$ have no channel state information (CSI) of the eavesdropper $E$. Then, the node $S$ cannot adapt the coding scheme to $E$'s channel state, but resorts to set the secrecy rate to a constant target rate $R_{s}$. When the instantaneous secrecy rate is larger than the target rate, i.e., $C_{s} > R_{s}$, perfect secrecy can be guaranteed. Otherwise, when the instantaneous secrecy rate is less than the target rate, i.e., $C_{s} \leq R_{s}$, secrecy will be compromised and secrecy outage occurs, the probability of which is given by the secrecy performance metric SOP \cite{barros2006secrecy}. The SOP is mathematically expressed as \cite{ai2020secrecy}
\begin{align}
P_{o}  =  &  \mathrm{Pr}\left[ C_{s}(  \gamma_{D}, \gamma_{E}  ) \leq  R_{s} \right]  =  \mathrm{Pr}\left[  \gamma_{D}  \leq \Theta \gamma_{E} + \Theta - 1  \right]   \nonumber  \\
       =  &  \int_{0}^{ \infty }  \int_{ 0 }^{ ( 1 + \gamma_{E} ) \Theta - 1 }  f_{ \gamma_{D}, \gamma_{E} }( \gamma_{D}, \gamma_{E} )  \,  d \gamma_{D}   d\gamma_{E},
\label{eq:sop_def}
\end{align}
where $\Theta=\exp(R_s) \geq 1$, and $f_{ \gamma_{D}, \gamma_{E} }(\cdot, \cdot)$ is the joint PDF of the RVs $\gamma_{D}$ and $\gamma_{E}$.

Next, we investigate the secrecy outage performance under V2V and V2I scenarios, respectively.


\subsection{V2V Communications \label{subsec:v2v_sop}}

With the RVs $\gamma_{D}$ and $\gamma_{E}$ being independent, and utilizing the Parseval's formula for Mellin's transform, the SOP can be alternatively evaluated as \cite{ai2019secrecyel}
\begin{align}
P_{o} 
      = & \int_{0}^{ \infty } \biggl[ \int_{ 0 }^{ ( 1 + \gamma_{E} ) \Theta - 1 }  f_{ \gamma_{D} }( \gamma_{D} )  \,  d \gamma_{D} \biggr] \cdot  f_{ \gamma_{E} }( \gamma_{E} )   d\gamma_{E}  \nonumber \\
      = &  \int_{\mathcal{L}_{1}} \!\!\!\! \mathcal{M}[F_{\gamma_{D}}( \Theta x \! + \! \Theta \! - \! 1 ), 1 \! - \! s] \cdot  \frac{ \mathcal{M}[f_{\gamma_{E}}(  x ), s] }{ 2 \pi j  }  \, ds ,
\label{eq:sop_def2}
\end{align}
where $\mathcal{L}_{1}$ is the integration path from $c-j\infty$ to $c+j\infty$ with $c$ being some constant, and $\mathcal{M}[f(x), s]$ represents the Mellin transform of the function $f(x)$ \cite[Eq.~17.41]{jeffrey2007table}.

\textbf{Lemma 1}: The SOP of the RIS-assisted vehicular communication under the V2V scenario as illustrated in Fig. \ref{fig:system_model} with perfect RIS phase shifting can be expressed as
\begin{align}
P_{o}   =   \mathcal{D}_{1}  \cdot   H^{0,1:1,1:1,2}_{1,0:2,2:1,2} \!\! \left( \!
\begin{array}{c}
\!\! \mathcal{E}_{1} \!\! \\
\!\! -  \!\!
\end{array}\middle\vert
\begin{array}{c}
\!\!  \mathcal{E}_{2}  \!\! \\
\!\!  \mathcal{F}_{2} \!\! \\
\end{array}\middle\vert
\begin{array}{c}
\!\!  \mathcal{E}_{3} \!\! \\
\!\!  \mathcal{F}_{3} \!\! \\
\end{array}\middle\vert
\frac{ \eta_{D} }{ \sqrt{\Theta  -  1 } }, \frac{ \Theta \overline{\gamma}_{E} }{ \Theta \! - \! 1 } \!
\right)  ,
\label{eq:v2v_sop2}
\end{align}
where  $\mathcal{D}_{1} \! = \!  \frac{  (\Theta  -   1)   }{ \overline{\gamma}_{E} \Theta \Gamma(k_{D}) } $, $\mathcal{E}_{1}  =  (2; 1, 1)$; $\mathcal{E}_{2}  =  ( 1 - k_{D}, 1), ( 1, 1)$; $\mathcal{F}_{2} =  (0, 1), (1, \frac{1}{2} )$; $\mathcal{E}_{3}  =  (1, 1), (1, 1)$; and $\mathcal{F}_{3} = (1, 1)$.

\emph{Proof}: Please refer to Appendix C\ref{app_sec:proposition3}.

\begin{figure*}[t!]
\centering
\begin{minipage}{.5\textwidth}
\centering
  \includegraphics[width=0.96\linewidth,keepaspectratio,angle=0]{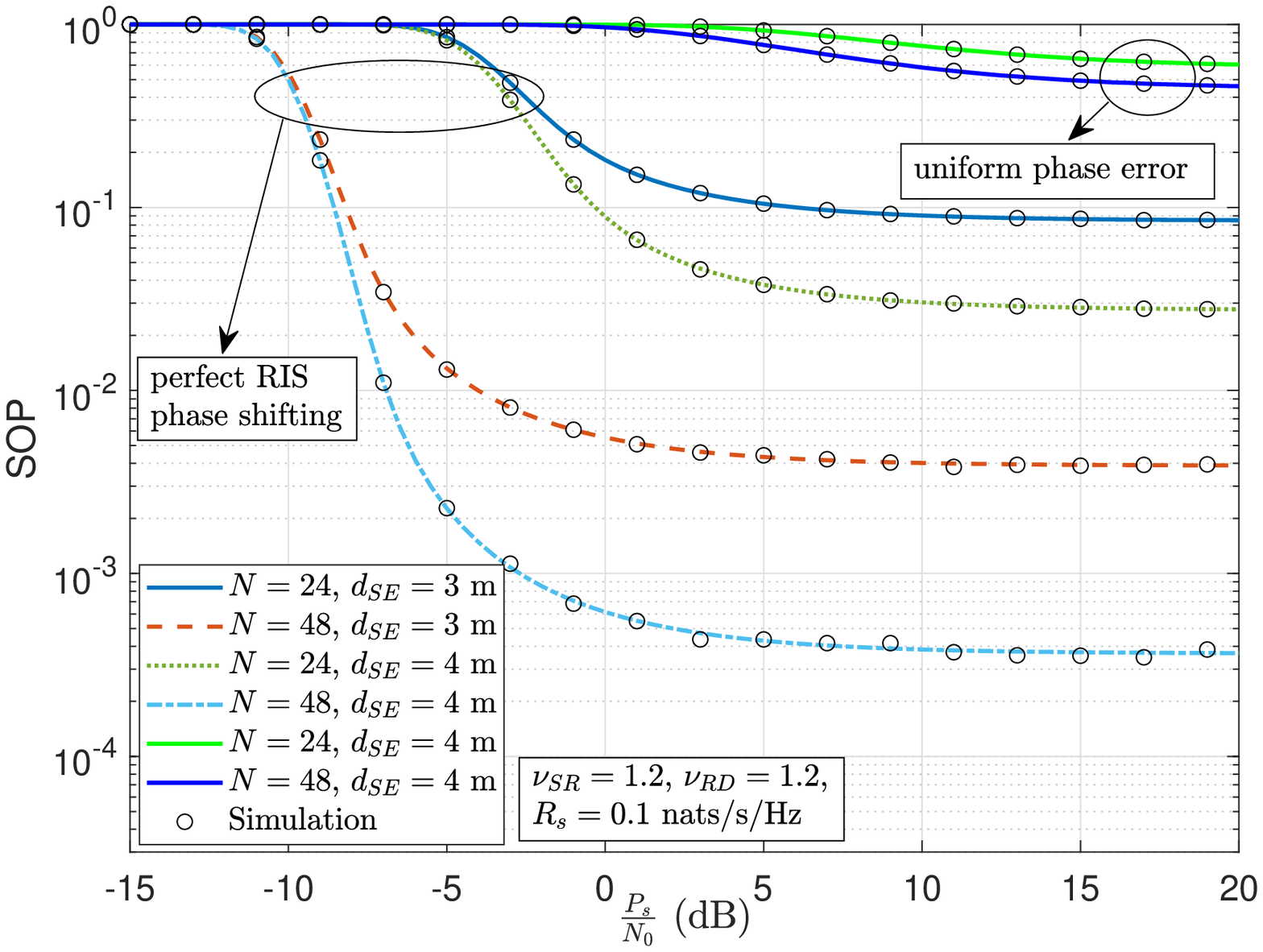}
  \caption{SOP vs. $\frac{P_{s}}{N_{0}}$ for varying number of RIS elements under V2V.}
  \label{fig:sop_v2v}
\end{minipage}\hfill
\begin{minipage}{.5\textwidth}
\centering
  \includegraphics[width=0.96\linewidth,keepaspectratio,angle=0]{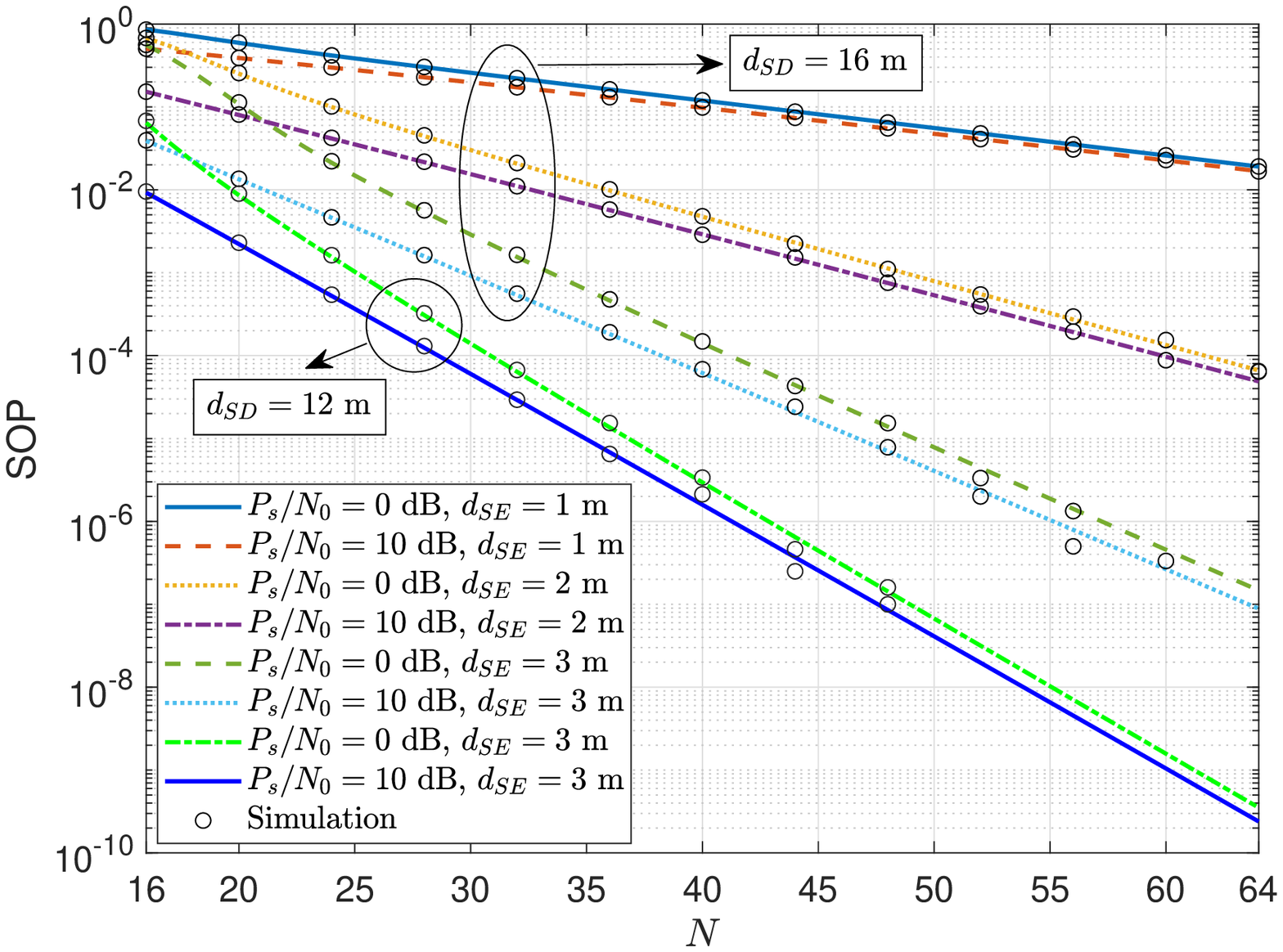}
  \caption{SOP vs. $N$ for varying number of RIS elements under V2I.}
  \label{fig:sop_v2i}
\end{minipage}\hfill
\end{figure*}

\textbf{Lemma 2}: The SOP of the RIS-assisted vehicular communication under the V2V scenario as shown in Fig. \ref{fig:system_model} with uniform distributed phase estimation error can be written as
\begin{align}
P_{o} \!  = \!  \mathcal{D}_{2}  \! \cdot \!  H^{0,1:1,2:1,2}_{1,0:2,3:1,2} \!\! \left( \!
\begin{array}{c}
\!\! \mathcal{E}_{4} \!\! \\
\!\! -  \!\!
\end{array}\middle\vert
\begin{array}{c}
\!\!  \mathcal{E}_{5}  \!\! \\
\!\!  \mathcal{F}_{5} \!\! \\
\end{array}\middle\vert
\begin{array}{c}
\!\!  \mathcal{E}_{6} \!\! \\
\!\!  \mathcal{F}_{6} \!\! \\
\end{array}\middle\vert
\frac{ \overline{\gamma}_{D} \nu_{SR}  }{ ( \Theta \!  - \!  1) \nu_{RD}^{-1}  }, \! \frac{ \Theta \overline{\gamma}_{E} }{ \Theta \! - \!\! 1 } \!
\right)  ,
\label{eq:v2v_sop3}
\end{align}
where  $\mathcal{D}_{2} =  (\mathcal{B} \Theta )^{-1}  ( \theta - 1 )^{ \frac{ N+ 3 }{ 2 } }  $, $\mathcal{E}_{4}  =  (\frac{N+5}{2}; 1, 1)$; $\mathcal{E}_{5}  =  ( \frac{3-N}{2}, 1), ( \frac{1+N}{2}, 1), ( \frac{3+N}{2}, 1)$; $\mathcal{F}_{5} =  (\frac{N-1}{2}, 1), (\frac{3+N}{2}, 1 )$; $\mathcal{E}_{6}  =  (1, 1), (1, 1)$; and $\mathcal{F}_{6} = (1, 1)$.

\emph{Proof}: The results follow by applying the same rationale as in Appendix C\ref{app_sec:proposition3}.


\subsection{V2I Communications \label{subsec:v2i_sop}}

Under the V2I scenario and with the independence between the RVs $\gamma_{D}$ and $\gamma_{E}$, the SOP can be evaluated as \cite{ai2019physicalpj}
\begin{align}
P_{o} = & \int_{0}^{\infty}    F_{\gamma_{D}}(\Theta x   +   \Theta   -   1 )   \cdot f_{\gamma_{E}}(x) \, dx .
\label{eq:v2i_sop1}
\end{align}

\textbf{Lemma 3}: The SOP of the RIS-aided vehicular network under the V2I scenario depicted in Fig. \ref{fig:system_model} can be evaluated by
\begin{align}
P_{o} = & 1 - \frac{ 1 }{ \overline{ \gamma }_{E} } \! \cdot \! \sum_{k = 1}^{N-1} \frac{ e^{ - \frac{ \Theta - 1 }{ \overline{\gamma}_{D} \Omega_{D} } } }{ ( \overline{\gamma}_{D} \Omega_{D} )^{k} \cdot k! }   \cdot \! \sum_{j=0}^{k}   \Theta^{k} \! \cdot \! {k \choose j} \!  \cdot \! \left( 1 - \frac{ 1 }{ \Theta } \right)^{k - j}  \nonumber \\
&  \cdot \left( \frac{ \Theta }{ \overline{\gamma}_{D} \Omega_{D} } \right)^{ - (j+1) } \cdot  G^{2,1}_{1,2}\!\left(  \frac{ \overline{ \gamma }_{D} \Omega_{D} }{ \Theta \overline{ \gamma }_{E} }   \left\vert^{ -j  }_{ 0, 0  } \right. \right) .
\label{eq:v2i_sop3}
\end{align}

\emph{Proof}: Please refer to Appendix D\ref{app_sec:proposition4}.


\section{Numerical Results and Discussions \label{sec:results}}

In this section, we numerically evaluate the secrecy outage performance under the considered V2V and V2I scenarios. The path loss exponents are set as $p_{1} = 2.1$ and $p_{2} = 2.3$ for simulation purpose.

Figure \ref{fig:sop_v2v} demonstrates the SOP in terms of the transmit SNR $\gamma_{t} = \frac{P_{s}}{N_{0}}$ under the V2V scenario. It can be seen that the SOP performance improves significantly by even a relatively small increase in the number of RIS antenna elements. However, when the SNR is large enough, the SOP performance stagnates and further increasing the transmit SNR (namely the transmit power) does not improve the SOP performance any longer. Instead, increasing the number $N$ can significantly improve this performance bound that can not be enhanced by increasing the transmission power.  It is obvious that the curves for perfect IRS shifting and random phase shifting represent the upper and lower limits of the secrecy outage performance, respectively, when only phase shifting is considered. The large differences between the lower and upper limits demonstrate the adverse effect of imperfect phase shifting for the RIS-aided system performance.

Figure \ref{fig:sop_v2i} shows the SOP performance improvement with the increase of the number of RIS elements $N$ under the V2I scenario. As expected, when the number of RIS elements $N$ increases, the SOP performance improves even when the signal attenuation for the legitimate receiver is much larger than that to the eavesdropper because of longer signal transmission. It is also observed that the SOP in the logarithm scale exhibits a linear relation with respect to the number $N$ when $N$ is large. The slope of the linear relation is dependent on the distances between the communicating entities and is irrelevant to the transmission power.

%


\begin{appendices}

\section*{Appendix A: Proof of Proposition 1 \label{app_sec:proposition1}}
We assume that the RV $\mathcal{A} = \sqrt{\overline{\gamma}_{D} } \cdot \sum_{n = 1}^{N} \alpha_{n} \beta_{n}$ in (\ref{eq:v2v_max_snr_d}) can be approximated by a Gamma RV $\mathcal{Z}$ with shape parameter $k_{D}$ and scale parameter $\eta_{D}$. It follows immediately that the first and second moments of the RV $\mathcal{Z}$ are $\mathbb{E}[\mathcal{Z}] = k_{D} \eta_{D}$ and $\mathbb{E}[\mathcal{Z}^{2}] = k_{D} \eta_{D}^{2}$, respectively. Since $h_{SR,n} \sim \mathcal{CN}(0, \nu_{SR} )$ and $\beta_{n}  \sim \mathcal{CN}(0, \nu_{RD} )$, we have that $\alpha_{n}=|h_{SR,n}|$ is a Rayleigh RV with $\mathbb{E}[\alpha_{n}] = \frac{\sqrt{\pi \nu_{SR}}}{2}$ and $\mathbb{E}[\alpha_{n}^{2}] = \nu_{SR}$. Similarly, we have $\mathbb{E}[\beta_{n}] = \frac{\sqrt{\pi \nu_{RD}}}{2}$ and $\mathbb{E}[\beta_{n}^{2}] = \nu_{RD}$. Next, we find the first and second moments of the RV $\mathcal{A}$ as follows:
\begin{align}
\mathbb{E}[\mathcal{A}]  = &  \sqrt{\overline{\gamma}_{D} } \! \cdot \! \sum\limits_{n = 1}^{N} \! \mathbb{E}\!\left[ \alpha_{n} \right] \mathbb{E}\!\left[ \beta_{n} \right]
=  \frac{ \sqrt{ \pi^{2} \nu_{SR} \nu_{RD} } }{ 4 \cdot ( \sqrt{\overline{\gamma}_{D} } \cdot N )^{-1} } , \\
\mathbb{E}[\mathcal{A}^{2}]  = & \mathbb{E}\!\biggl[\!\!\biggl(\sqrt{\overline{\gamma}_{D} } \cdot \sum\limits_{n = 1}^{N} \alpha_{n} \beta_{n}\biggr)^{2}\biggr] \! = \! \overline{\gamma}_{D} \! \cdot \! \sum\limits_{n = 1}^{N} \mathbb{E}\!\left[ \alpha_{n}^{2} \right] \mathbb{E}\!\left[ \beta_{n}^{2} \right] \nonumber  \\  &  +  \overline{\gamma}_{D} \cdot \sum\limits_{n = 1}^{N}\sum\limits_{m \neq n }^{N} \mathbb{E}\!\left[ \alpha_{n} \right]  \cdot \mathbb{E}\!\left[ \beta_{n} \right] \mathbb{E}\!\left[ \alpha_{m} \right]  \mathbb{E}\!\left[ \beta_{m} \right] \nonumber \\
= & \overline{\gamma}_{D} \cdot \!\! \left[  N  \nu_{SR}  \nu_{RD} +  \frac{   \pi^{2} \nu_{SR} \nu_{RD}  \cdot N (N-1) }{ 16 } \right] .
\label{eq:v2v_statistics_aux2b}
\end{align}

%
%

Solving equalities $\mathbb{E}[\mathcal{A}] = \mathbb{E}[\mathcal{Z}]$ and $\mathbb{E}[\mathcal{A}^{2}] = \mathbb{E}[\mathcal{Z}^{2}]$, we obtain the values of $k_{D}$ and $\eta_{D}$, which determines the statistics of Gamma RV $\mathcal{Z}$. Finally, with change of RV $\gamma_{D} = \mathcal{Z}^{2}$, we obtain the statistic function of RV $\gamma_{D}$ as in Proposition 1.

\section*{Appendix B: Proof of Proposition 2 \label{app_sec:proposition2}}

The RV $r = \bigl|  \sum_{n = 1}^{N} \alpha_{n} \beta_{n} e^{j \sigma_{n}^{D}}   \bigr|$ can be interpreted as the distance to origin after $N$ 'random walks' with the $n$-th step of length $r_{n} = \alpha_{n} \beta_{n}$ in the direction of $\sigma_{n}^{D}$. We first obtain the statistics of the RV $r_{n}$, which is the product of two i.n.i.d Rayleigh RVs.
\begin{align}
f_{r_{n}}(r)  = & \int_{0}^{\infty} \! \frac{1}{x}  f_{\alpha_{n}}\!(x) f_{\beta_{n}}\!\!\left( \frac{r}{x} \right) dx
          \! = \!\!\!   \int_{0}^{\infty} \! \frac{ 4r \! \cdot \! e^{ - \frac{x^{2}}{ \nu_{SR} } - \frac{ r^{2} }{ \nu_{RD} x^{2} }  } }{ x \nu_{SR} \nu_{RD} } dx \nonumber \\
          = & \frac{ 4r }{ \nu_{SR} \nu_{RD} } \cdot K_{0}\!\!\left( \frac{ 2r }{ \nu_{SR} \nu_{RD} } \right) .
\label{eq:v2v_pdf_snr_aux1}
\end{align}


Since the angle density functions of $\sigma_{n}^{D}$ are uniform, the conditional PDF of RV $r$ is given with the Kluyver's result in terms of integral over Bessel functions \cite[p.~420]{watson1995treatise}
\begin{align}
f_{r}(r|r_{1},\cdots, r_{N}) \! = \!\!\! \int_{0}^{\infty} \! \frac{ r J_{0}\!(rx)}{ x^{-1} }  J_{0}\!(r_{1}x) \dots J_{0}\!(r_{N}x)   dx  .
\label{eq:v2v_pdf_snr_aux2}
\end{align}

Then, the PDF of the RV $r$ can be obtained directly from
\begin{align}
& f_{r}(r)  =  r  \left[ \frac{ 4 }{ \nu_{SR} \nu_{RD} } \right]^{N} \!\!\! \int_{0}^{\infty}  \!\!\!\! x J_{0}(rx)  \!\! \int_{0}^{\infty} \!\!\!\!\! \cdots \!\!\! \int_{0}^{\infty} \!\!\! r_{1} K_{0}\!\biggl(\! \frac{ 2r_{1} }{ \nu_{SR} \nu_{RD} } \!\biggr) \nonumber \\ & \quad \cdot \! J_{0}( r_{1}x ) dr_{1}   \dots  r_{N} K_{0}\!\biggl(\! \frac{ 2r_{N} }{ \nu_{SR} \nu_{RD} } \! \biggr) \cdot J_{0}( r_{N}x )  dr_{N}   dx  .
\label{eq:v2v_pdf_snr_aux3}
\end{align}

Next, utilizing the following equalities \cite[Chpt.~13]{watson1995treatise}
\begin{align}
& \int_{0}^{\infty} \!\!\! r_{n} K_{0}\!\!\left( \! \frac{ 2r_{n} }{ \nu_{SR} \nu_{RD} } \! \right)  J_{0}( r_{n}x ) d r_{n} \! = \! \frac{ ( \nu_{SR} \nu_{RD} )^{2} }{ (\nu_{SR} \nu_{RD} x )^{2} \! + \! 4 },  \\
& \int_{0}^{\infty} \!\!  \frac{ ( \nu_{SR} \nu_{RD} )^{2N}  J_{0}(rx) x }{ [ (\nu_{SR} \nu_{RD} x )^{2} \! + \! 4 ]^{N} }  dx \! = \!  \frac{ 4^{ 1  -  N}  K_{ N-1 }\!\bigl(\! \frac{ 2r }{ \nu_{SR} \nu_{RD} } \!\bigr)  }{ \Gamma\!(N)   (\nu_{SR} \nu_{RD} r)^{ 1 - N}  } ,
\label{eq:v2v_pdf_snr_aux4}
\end{align}
in (\ref{eq:v2v_pdf_snr_aux3}) leads to the PDF of the RV $r$. With change of RV $\gamma_{D} = \overline{\gamma}_{D}r^{2}$, we obtain the exact PDF of the RV $\gamma_{D}$ in (\ref{eq:v2v_pdf_snr_d2}). The CDF of $\gamma_{D}$ in (\ref{eq:v2v_cdf_snr_d2}) follows by using its relation with the PDF.

\section*{Appendix C: Proof of Lemma 1 \label{app_sec:proposition3}}
We first solve the function $\mathcal{M}[F_{\gamma_{D}}( \Theta x + \Theta - 1 ), 1 - s]$. Based on the definition of Mellin transform, we have
\begin{align}
& \mathcal{M}[F_{\gamma_{D}}( \Theta x  +   \Theta  -  1 ), 1  -  s] \!  = \! \int_{0}^{\infty} \!\!\!  x^{-s}  F_{\gamma_{D}}( \Theta x  +  \Theta  -  1 ) \, dx  \nonumber \\
& \stackrel{(a)}{=} \int_{0}^{\infty} \!  \frac{ x^{-s} }{ \Gamma( k_{D} ) } \cdot  \MeijerG*{1}{1}{1}{2}{\! 1 \! }{\! k_{D}, 0 \! }{\! \frac{ \sqrt{ \Theta x + \Theta - 1 } }{ \eta_{D} } }  \, dx   \stackrel{(b)}{=} \frac{ ( 2 \pi j )^{-1} }{  \Gamma( k_{D} )   } \nonumber \\
&  \quad \cdot \int_{\mathcal{L}_{2}} \!\! \int_{0}^{\infty} \!\!\!\! x^{-s}  \frac{ \Gamma( k_{D} \! + \! \xi ) \Gamma( - \xi ) }{ \Gamma( 1 \! - \! \xi ) } \!\! \left[ \frac{ \sqrt{ \Theta x \! + \! \Theta \! - \! 1 } }{ \eta_{D} } \right]^{ - \xi } \!\!\! dx d\xi,
\label{eq:mellin1}
\end{align}
where $(a)$ is obtained by converting the incomplete Gamma function in (\ref{eq:v2v_cdf_max_snr_d}) into Meijer G-function; and $(b)$ is derived by representing the Meijer G-function in terms of  contour integral and changing the integration order. The inner integral in (\ref{eq:mellin1}) can be solved with the help of \cite[Eq.~(3.194.3)]{jeffrey2007table} as
\begin{align}
\int_{0}^{\infty} \! \frac{ x^{-s} }{ ( \Theta x  +  \Theta  -  1 )^{ \frac{\xi}{2}} }  \,dx  =  \frac{ \Gamma( 1  -  s ) \Gamma( s  +  \frac{\xi}{2}  -  1 ) }{ \Gamma( \frac{\xi}{2} ) ( \Theta  -  1 )^{ \frac{\xi}{2} }  \bigl( \frac{ \Theta }{ \Theta  -  1 }  \bigr)^{1-s}  }   .
\label{eq:mellin1_app}
\end{align}

Substituting (\ref{eq:mellin1_app}) into (\ref{eq:mellin1}), we obtain after some algebra
\begin{align}
& \mathcal{M}[F_{\gamma_{D}}( \Theta x \! + \!  \Theta \! - \!  1 ), 1 \! - \! s] \stackrel{(c)}{=} \frac{ \Gamma( 1 \! - \! s ) }{ \Gamma( k_{D} ) } \cdot \left( \frac{ \Theta }{ \Theta  \! - \! 1 } \right)^{ s - 1 } \cdot \frac{ 1 }{ 2 \pi j } \nonumber \\
& \quad\!\!   \cdot  \int_{\mathcal{L}_{2}} \!\!\! \frac{ \Gamma( k_{D} \! + \! \xi ) \Gamma( - \xi ) \Gamma( s \! + \! \frac{ \xi }{ 2 } \! - \! 1 ) }{ \Gamma( 1 \! - \! \xi ) \Gamma( \frac{ \xi }{ 2 } ) } \cdot \left( \frac{ \sqrt{ \Theta \! - \! 1 } }{ \eta_{D} } \right)^{ - \xi } \, d\xi ,
\label{eq:mellin2a}
\end{align}
where $(c)$ is obtained with the aid of \cite[Eq.~(3.194.3)]{jeffrey2007table} with $\mathcal{L}_{2}$ being some contour.


Rewriting the Bessel function in (\ref{eq:v2i_pdf_snr_e}) in Meijer G-function's form, the Mellin transform $\mathcal{M}[f_{\gamma_{E}}(  x ), s]$ can be solved as
\begin{align}
\mathcal{M}[f_{\gamma_{E}}(  x ), s] \! = \!\! \int_{0}^{\infty} \!\! \frac{ x^{ s-1 } }{ \overline{\gamma}_{E} }  \cdot  G^{2,0}_{0,2}\!\left( \!\! \frac{ x }{  \overline{ \gamma }_{E} }   \left\vert^{ -  }_{ 0, 0  } \right. \!\! \right)  dx
\! = \!  \frac{ \Gamma\!\left( s \right)  \cdot  \Gamma\!\left( s \right)  \overline{\gamma}_{E}^{s}  }{ \overline{\gamma}_{E} } .
\label{eq:mellin3}
\end{align}

Next, substituting (\ref{eq:mellin2a}) and (\ref{eq:mellin3}) into (\ref{eq:sop_def2}), the SOP can be expressed, after some mathematical manipulations, as the following double contour integral:
\begin{align}
P_{o}   = &  \frac{  (\Theta  -  1)   }{ \overline{\gamma}_{E} \Theta \Gamma(k_{D}) } \cdot \left(\frac{1}{2 \pi j}\right)^{2}   \cdot  \int_{\mathcal{L}_{1}} \!\! \int_{\mathcal{L}_{2}}  \!\!  \left( \frac{ \eta_{D} }{ \sqrt{ \Theta  -  1 } } \right)^{\xi}  \nonumber \\ & \cdot  \left( \frac{  \Theta \cdot \overline{\gamma}_{E} }{\Theta-1} \right)^{s} \cdot \frac{ \Gamma( 1 - s ) \cdot \Gamma( s ) \cdot \Gamma( s ) }{ \Gamma(1-\xi) \cdot \Gamma(\frac{\xi}{2}) }  \cdot \Gamma\!\left( k_{D} + \xi \right)  \nonumber \\ &  \cdot \Gamma\!\left( - \xi \right) \cdot \Gamma\!\left( s + \xi - 1 \right)  d\xi ds .
\label{eq:v2v_sop1}
\end{align}

Recalling the representation of bivariate Fox H-function in terms of double contour integral \cite[Eq.~2.56]{mathai2009h} for (\ref{eq:v2v_sop1}), the exact expression for the SOP can be expressed in terms of bivariate Fox H-function as shown in Lemma 1.


\section*{Appendix D: Proof of Lemma 2 \label{app_sec:proposition4}}
Expressing the incomplete Gamma function in (\ref{eq:v2i_cdf_snr_d}) in series \cite[Eq.~(8.354)]{jeffrey2007table} and then substituting the resulting expression of CDF and (\ref{eq:v2i_pdf_snr_e}) into (\ref{eq:v2i_sop1}), the SOP can be rewritten as

\begin{align}
P_{o} \! = & \!\! \int_{0}^{\infty} \!\!\! \frac{ 2 }{ \overline{\gamma}_{E} }   K_{0}\!\biggl( 2 \sqrt{  \frac{ x }{ \overline{\gamma}_{E} }  } \biggr) \!\! \cdot \!\! \biggl[ \! 1 \! - \! e^{ - \frac{ ( 1 \! + \! x ) \Theta \! - \! 1 }{ \overline{\gamma}_{D} \Omega_{D} } } \!\! \cdot \!\!\! \sum_{k=1}^{N \!- \! 1} \!\! \frac{ [ ( 1 \! + \! x )\Theta \! - \! 1 ]^{k} }{ ( \overline{\gamma}_{D} \Omega_{D} )^{k} \cdot k! } \! \biggr]   dx \nonumber \\
= & 1 - \frac{ 2 }{ \overline{ \gamma }_{E} } \cdot e^{ - \frac{ \Theta - 1 }{ \overline{\gamma}_{D} \Omega_{D} } } \cdot \sum_{k = 1}^{N-1} \frac{ 1 }{ ( \overline{\gamma}_{D} \Omega_{D} )^{k} \cdot k! }  \nonumber \\
& \cdot \!\! \int_{0}^{\infty} \!\! K_{0}\!\biggl( 2 \sqrt{  \frac{ x }{ \overline{\gamma}_{E} }  } \biggr) \! \cdot  e^{ - \frac{ \Theta x }{ \overline{\gamma}_{D} \Omega_{D} } } \! \cdot \left[ (1 \! + \! x)\Theta - 1 \right]^{k}  dx .
\label{eq:v2i_sop2}
\end{align}

Rewriting $\left[ (1 + x)\Theta - 1 \right]^{k}$ in series and relevant terms in Meijer G-functions, the definite integral in (\ref{eq:v2i_sop2}) becomes
\begin{align}
\mathcal{I} = &  \sum_{j=0}^{k} \frac{ \Theta^{k} \cdot {k \choose j} }{ \left( 1 \!\! - \!\! \frac{ 1 }{ \Theta } \right)^{ j - k }  } \! \cdot \!\!\! \int_{ 0 }^{ \infty } \!\!\! x^{j}  \! \cdot \! G^{1,0}_{0,1}\!\biggl( \!\! \frac{ \Theta x }{ \overline{ \gamma }_{D} \Omega_{D} }   \left\vert^{ - }_{ 0  } \right. \!\! \biggr) \! \cdot \!  G^{2,0}_{0,2}\!\biggl(\!  \frac{ x }{ \overline{ \gamma }_{E}  }   \left\vert^{ - }_{ 0 , 0 } \right. \!\! \biggr) \!  dx \nonumber \\
= & \sum_{j=0}^{k} \frac{ \Theta^{k}  \cdot {k \choose j} \cdot \left( 1 - \frac{ 1 }{ \Theta } \right)^{k - j} }{ 2 \cdot \bigl( \frac{ \Theta }{ \overline{\gamma}_{D} \Omega_{D} } \bigr)^{ (j+1) } }  \cdot  G^{2,1}_{1,2}\!\biggl(  \frac{ \overline{ \gamma }_{D} \Omega_{D} }{ \Theta \overline{ \gamma }_{E} }   \left\vert^{ -j  }_{ 0, 0  } \right. \biggr) .
\label{eq:v2i_sop2_app1}
\end{align}


Finally, substituting (\ref{eq:v2i_sop2_app1}) into (\ref{eq:v2i_sop2}) leads to the closed-form expression for the SOP under the V2I scenario in Lemma 2.

\end{appendices}

%

\ifCLASSOPTIONcaptionsoff
  \newpage
\fi


\balance
\bibliographystyle{IEEEtran}
\bibliography{PLS_ris_aided_v2v_ref}

%
%


\end{document}